\documentclass[twocolumn,british,aps,prl,nofootinbib,floatfix]{revtex4-2}
\usepackage{mathpazo}
\usepackage{helvet}
\usepackage{courier}

\usepackage[T1]{fontenc}
\usepackage[utf8]{inputenc} %
\usepackage[a4paper]{geometry}
\usepackage{babel}
\usepackage{units}
\usepackage{amsmath} %
\usepackage{amsthm} %
\usepackage{amssymb}
\usepackage[unicode=true]
 {hyperref}

\makeatletter
\numberwithin{figure}{section}

\usepackage{graphicx}
\usepackage{palatino}
\usepackage{scalerel}
\usepackage{mathtools} %
\usepackage{tikz}
\hypersetup{colorlinks=true,
            linkcolor=blue,
            filecolor=magenta,      
            urlcolor=cyan,
			breaklinks=true
           } %

\usepackage{contour}
 \usepackage[normalem]{ulem}

\contourlength{0.8pt}

\newcommand{\tightoverset}[2]{%
  \mathop{#2}\limits^{\vbox to 0.ex{\kern-1ex\hbox{$#1$}\vss}}}

\makeatother

\usepackage{mathtools}
\usepackage{csquotes}
\usepackage{physics}
\usepackage{nicefrac}
\setcitestyle{square}

\usepackage{xcolor}
\definecolor{modification}{rgb}{0, 0, 0}
\definecolor{movedtext}{rgb}{0, 0, 0}
\definecolor{furthermodification}{rgb}{0, 0, 0}

\global\long\def\ket#1{|#1\rangle}%
 
\global\long\def\bra#1{\langle#1|}%

\global\long\def\bk#1#2{\langle#1|#2\rangle}%
 
\global\long\def\kb#1#2{|#1\rangle\langle#2|}%

\global\long\def\psip{\psi}

\global\long\def\rhop{\rho}

\begin{document}

\title{A Quantum Theory with Non-Collapsing Measurements} 
\author{Vincenzo Fiorentino and Stefan Weigert}
\date{August 2025}
\affiliation{Department of Mathematics, University of York~\\
York YO10 5GH, United Kingdom~\\
vincenzo.fiorentino@york.ac.uk, stefan.weigert@york.ac.uk~\\
}
\begin{abstract}
A collapse-free version of quantum theory is examined to systematically study the role of the projection postulate. This foil theory assumes ``passive'' measurements that do not update quantum states although measurement outcomes still occur probabilistically, and in accordance with Born's rule. The Hilbert space setting of quantum theory is retained. ``Passive quantum theory'' is shown to reproduce preparational uncertainty relations, the no-cloning theorem, and no-signalling, among other properties. Striking differences occur, however, if protocols involve post-measurement states. For example, a \textit{single} system, rather than an ensemble, is sufficient to reconstruct the state of the system. The possibility to ``observe'' a state increases the computational power of some quantum algorithms. Passive quantum theory is not locally tomographic but capable  of ``simulating'' quantum measurements modulo a finite delay. Outcome probabilities for composite systems may violate Bell inequalities, without however entailing an argument against local hidden variables.
\end{abstract}
\maketitle

\section{Motivation}
Experimental evidence for the state update of a quantum system induced by measurements was available since 1925. Using a cloud chamber, Compton and Simon \cite{compton_directed_1925} studied the scattering of ``x-ray quanta'' by electrons. They discovered that the angle characterizing the path of the recoiling electron and the angle of the photon scattering direction are strongly corre-lated. Knowing one of them is sufficient to determine where the particles interacted. These ``position measurements'' can be carried out in arbitrary temporal order.

According to von Neumann \cite{von_neumann_mathematische_1971},  this experiment implements two subsequent measurements of one single observable, namely the spatial coordinate of the interaction locus. The measurements of the angles can happen in quick succession and lead to identical re-sults. The
resulting \textit{deterministic repeatability} is argued by von Neumann to be equivalent to assuming the \textit{projection postulate}: immediately after measuring a non-degenerate observable, a quantum system will reside in the unique eigenstate associated with the observed outcome. In principle, Nature could have decided to realize other relations between the outcomes of two identical  consecutive measurements in quick succession. Von Neumann mentions two options. Either a 
\textit{deterministic} mechanism could control the measurement outcomes (this assumption effectively amounts to the existence of hidden variables), or the outcomes of the second measurement could be governed by the same probability distribution as the outcomes of the first. 

In this paper, we investigate a quantum-like theory that realizes von Neumann's second option: measuring an observable causes \textit{no} update of the state of the system. All other features of quantum theory, such as its setting in Hilbert space and the Born rule, are retained. The resulting \textit{passive} quantum theory (pQT) shares many features with standard quantum theory but is manifestly different from it. As a foil theory, pQT does not aim to reproduce quantum theory, unlike unitary models that try to eliminate the projection postulate. 

One of our main motivations is to examine the role of the projection postulate in quantum theory: it should become evident which properties cease to hold when the collapse is suspended. \textcolor{furthermodification}{This} scenario has not yet been investigated systematically; Sec.\ \ref{Sec: other work} briefly summarizes earlier work relevant to our approach. Our study is, in fact, a first step towards investiga-ting operationally consistent, quantum-like theories in which the familiar state-update rule of quantum theory is replaced by another rule \cite{fiorentino_quantum_nodate}. 

\section{Passive quantum theory} \label{sec: pQT axioms}
Five axioms define a bare-bones version of (non-relativistic) quantum theory. Four of these axioms describe
the mathematical framework of the theory: (${\cal S}$) the \emph{states} of a quantum system correspond to rays $\ket{\psi}$ in a separable, complex Hilbert space ${\cal H}$ and to their probabilistic mixtures $\rho$; (${\cal O}$) \emph{observables} are represented by Hermitean operators $\hat{A}$ acting on the space ${\cal H}$; (${\cal T}$) the \emph{time evolution} of quantum states is governed by Schrödinger's equation; (${\cal C}$) the state space of a \emph{compo-site system} is obtained by tensoring the Hilbert spaces of its constituent parts.

The fifth axiom ${\cal M}$ relates theory to experiment. Its three parts specify (${\cal M}{}_{1}$) the \emph{measurement outcomes }(the
eigenvalues $a_{r}$ of the measured observable $\hat{A}$), (${\cal M}_{2}$) the \emph{probability }with which they will occur\footnote{\textcolor{furthermodification}{We emphasise that Axiom ${\cal M}_{2}$ is understood to assign outcome probabilities to each \textit{individual} measurement; on its own, it does not account for \textit{correlations} between outcomes of \textit{multiple} measurements carried out sequentially or by distinct parties in compo-site systems.
This distinction proves significant when examining modifications of ${\cal M}_{3}$ in Sec.\ \ref{sec: conclusions}.}} (Born's rule:
$p_{A}(a_{r})=\left|\bk{a_{r}}{\psi}\right|^{2}$ for non-degenerate $\hat{A}$), and (${\cal M}_{3}$) the \emph{post-measurement states} (the eigenstates $\ket{a_{r}}$ of $\hat{A}$ when $a_{r}$ has been observed). As a mapping, the projection postulate ${\cal M}_{3}$ states that
\begin{equation}
\ket{\psi} \;\; \stackrel{a_{r}}{\longrightarrow} \;\; \hat{P}_{r}\ket{\psi}/\sqrt{\bra{\psi}\hat{P}_{r}\ket{\psi}}\,,\label{eq: M_3 in QM}
\end{equation}
where $\hat{P}_{r}=\kb{a_{r}}{a_{r}}$ projects onto the state $\ket{a_{r}}$. The quantum states undergo \emph{non-linear} transformations.

The model pQT is defined by the same set of postulates as quantum theory, except for the state-update rule ${\cal M}_{3}$. It will be convenient  to refer to states and measurements in pQT as \emph{p-states} and \textit{p-measurements} etc.
The modified projection postulate reads\vspace{-1mm}

\begin{quotation}
\hspace{-1.2cm}$\mathcal{M}^{\prime}_3$: A system resides in the same p-state $\ket{\psip}$\emph{ before and after}
measuring an observable $\hat{A}$.
\end{quotation}
Thus, p-states undergo a \emph{linear} transformation when measurements are carried out,
\begin{equation}
\ket{\psip} \, \stackrel{a_{r}}{\longrightarrow} \, \ket{\psip}\,.\label{eq: M_3 in pQT}
\end{equation}
Our main goal is to investigate the consequences of replacing rule \eqref{eq: M_3 in QM} by \eqref{eq: M_3 in pQT}.

\section{Single p-systems }
\subsection{Uncertainty relations}
The predictions of standard quantum theory and pQT agree as long as post-measurement states are neither used nor referred to. The expectation value of an observable $\hat{A}$ in pQT, for example, can be determined just as in quantum theory: the eigenvalue $a_{r}$ of the observable $\hat{A}$ will occur with probability $p_{A}(a_{r})$ upon measuring it repeatedly on an \emph{ensemble} of systems each of which resides in the p-state $\ket{\psip}$.
Hence, preparational uncertainty relations \cite{kennard_zur_1927,schrodinger_about_1930}
for the variances of non-commuting observables do hold in pQT. It follows that the inequalities exist solely due to the probabilistic character of measurement outcomes---state changes caused by quantum measurements cannot be their source. \textcolor{modification}{This property of pQT represents an independent argument against any disturbance-based interpretation of preparational uncertainty relations in quantum theory.} 

\textcolor{movedtext}{In other words, Heisenberg's original plausibility argument---measuring the position of an electron will cause an uncertainty of its momentum, due to an uncontrollable state change \cite{heisenberg_uber_1927}---is invalid for preparational uncertainties. Heisenberg's classically inspired reasoning needs to be justified in terms of \emph{measurement} uncertainty \cite{busch_heisenbergs_2007,busch_measurement_2014}.}

\textcolor{modification}{Replacing projective measurements by passive ones also suppresses the well-known  \textit{dual} role of quantum mechanical measurements. On the one hand, their outcomes provide information about the measured state; on the other hand, they will, generally, leave the system in a different, possibly pure state. In contrast, passive measurements cannot be used to ``prepare'' states in this sense, and, \textit{a fortiori}, will also not ``disentangle'' an entangled state of a multipartite system (cf.\ Sec.\ \ref{jointprobs}).}    

It becomes obvious that pQT will deviate from quantum theory whenever  post-measurement states play a role. For example, the expectation value of an observable may be obtained in pQT from a \emph{single} copy of a p-state $\ket{\psip}$, in contrast to the ensemble needed in quantum theory. Since a (non-destructive) passive measurement of $\hat{A}$ does not update a p-state, it is possible to repeat it \emph{on one and the same }system as often as is necessary to determine the outcome probabilities $ p{}_{A}(a_{r})=\bra{\psi}\hat{P}_{r}\ket{\psi}$. \textcolor{modification}{This observation has far-reaching consequences.}

\subsection{Single-copy state reconstruction }
Both a quantum state and a p-state correspond to a unique ray in the Hilbert space of the system. However, they differ from an operational point of view: the collapse-free theory allows us to reconstruct an unknown p-state $\ket{\psip}$ from a \emph{single} system \cite{busch_is_1997, kent_nonlinearity_2005, aaronson_space_2016}. To identify the ray in Hilbert space associated with $\ket{\psip}$, one simply repeats p-measurements of an informationally complete set
of observables \cite{prugovecki_information-theoretical_1977,busch_informationally_1991,ivanovic_geometrical_1981}
on the given system, without the need for an ensemble.

What is more, an experimenter can tell apart any two distinct non-orthogonal p-states $\ket{\psip}$ and $\ket{\phi}$ \emph{with
certainty}, even when being presented with a single copy only. Successful state reconstruction and discrimination based on a single copy of the unknown state means that p-states represent observable, \emph{classical} quantities. From an ontological point of view, this ``reality'' of p-states in pQT removes notorious interpretational issues posed by quantum states.

The absence of state updates after p-measurements means that only a \emph{single }dynamical law exists in pQT, described by Axiom ${\cal T}$. Hence, the tension between the unitary dynamics of a quantum system and its ``stochastic'' time evolution caused by measurements is entirely absent in pQT. Attempts to elimi-nate the non-deterministic evolution from quantum theory have a long history, ranging from models which consider the measurement device as a quantum system \cite{von_neumann_mathematische_1971,arthurs_simultaneous_1965} to alternative interpretations of the theory \cite{dewitt_many-worlds_2015,dieks_modal_1998}. In pQT, this problem does not arise. Nevertheless, part of the measurement problem persists in the sense that Axiom ${\cal T}$ appears insufficient
to explain the emergence of specific measurement outcomes.

\subsection{Density operators }
Gleason's theorem \cite{gleason_measures_1975}
tells us that mixed states emerge naturally in the Hilbert space setting
of quantum theory. The proof of the theorem, based on associating
measurement outcomes with projection operators, remains valid in pQT.
Thus, non-negative Hermitean operators with unit trace also represent
candidates for states in pQT. However, their interpre-tation as \emph{mixtures}
of pure quantum states cannot be upheld in pQT as it is possible to
identify an unknown p-state $\ket{\psip}$ by carrying
out single-copy state reconstruction, at least in principle \cite{busch_is_1997}. If the
ignorance contained in the classical probability of a p-density operator
can always be removed, \emph{proper} (or epistemic) mixtures do not
represent a fundamental concept in pQT.

However,\emph{ improper} p-density matrices still play a role in pQT.
They arise if an observer can access only a part of a larger p-system, just as in quantum theory. To see this, we need to discuss the behaviour of composite systems when p-measurements are carried out. For simplicity, we will limit ourselves to bipartite systems.

\section{Composite p-systems}
\subsection{Measurements}
The pure states $\ket{\Phi}$ of a bipartite quantum system are elements of the space ${\cal H}_{AB}={\cal H}_{A}\otimes{\cal H}_{B}$
(cf.\ Axiom ${\cal C}$). The corresponding bipartite p-system has the same state space. The mathematical distinction between product states and entangled states in the space ${\cal H}_{AB}$ applies equally to pQT and quantum theory. 

An experimenter who has access to a part of a composite system only can still measure ``local'' observables, using a ``local'' device. This situation is characteristic for Bell-type experiments. Having observed the outcome $a_{r}$ upon measuring the observable $\hat{A}$,  a suitable projection operator describes the update of the initial \emph{quantum} \textit{state} $\ket{\Phi}$,
\begin{equation}
\ket{\Phi}\, \stackrel{a_{r}}{\longrightarrow}\, \hat{P}_{r}\otimes\mathbb{I}\ket{\Phi}/\sqrt{\bra{\Phi}\hat{P}_{r}\otimes\mathbb{I}\ket{\Phi}}\,.\label{eq: M_3 in QM bipartite}
\end{equation}

In contrast, a p-measurement of the local observable $\hat{A}$  does \emph{not} cause the \textit{p-state} $\ket{\Phi}$ to update if we 
extend the map  \eqref{eq: M_3 in pQT}  to states of a composite system and arbi-trary measurements performed on them. In quantum theory, the probability to obtain the value $a_{r}$ is found from repeated measurements on an \textit{ensemble} of systems in state $\ket{\Phi}$, 
\begin{equation} \label{eq: p_a probability}
    p_{A}(a_{r})=\bra{\Phi}\hat{P}_{r}\otimes\mathbb{I}\ket{\Phi}\,.
\end{equation}
In pQT the value of the probability $ p_{A}(a_{r})$ can be found from repeated measurements of $\hat{A}$ on a \textit{single} system. Quantum theory and its foil predict the same numerical value $ p_{A}(a_{r})$. 

Interestingly, given a single physical system, local p-measurements on a subsystem can reveal whether 
the composite system resides in a product state $\ket{\phi_{A}}\otimes\ket{\phi_{B}}$ or in an entangled state $\ket{\Phi}$: single-copy state
reconstruction will return either the p-state $\ket{\phi_{A}}$ or the \emph{mixed} p-state $\rhop_{A}=\text{Tr}_{B}\ket{\Phi}\bra{\Phi}$,
respectively. In this case, the use of a p-density matrix $\rhop_{A}$ is both necessary and justified since it provides the correct description of the subsystem as seen by a local observer.

\subsection{Joint probabilities}\label{jointprobs}
Passive measurements on subsystems do not create correlations between entangled systems since they do not collapse
p-states. Therefore, \emph{local }p-measurements cannot reveal the
probabilities of \emph{joint} outcomes which are essential to probe
Bell-type inequalities. \emph{Global} p-measurements must be used
to extract joint outcome probabilities which may violate Bell-type
inequalities. However, a violation would \emph{not} imply the existence
of non-classical correlations and, hence, could not be used to rule out the existence
of local hidden variables. It is also impossible to reconstruct entangled
p-states from local p-measurements which means that pQT is not
locally tomographic.

\section{Quantum information with passive measurements }
\subsection{Cloning p-states}
The no-cloning theorem \cite{wootters_single_1982,dieks_communication_1982}
states that it is impossible to \textit{dynamically} generate copies of an
unknown quantum state, i.e.\ through the application of unitary gates.
This result also holds in pQT since no quantum measurements are required to derive it. However, an alternative cloning procedure exists which requires a single system only. Once the state of a p-system has been revealed by single-copy state-reconstruction,
another system can be prepared in the observed p-state. In quantum
theory, such a measurement-based copying procedure would require an
\emph{ensemble} of identically prepared systems.

\subsection{Teleportation, quantum cryptography and quantum computation}
The state update induced by quantum measurements is essential for
many protocols of quantum information. Teleportation \cite{bennett_teleporting_1993}
and entanglement swapping \cite{pan_experimental_1998}, for example,
rely on system-wide state changes as a result of local measurements.
Thus, they will no longer work in pQT. The impossibility to ``steer''
the state of a distant subsystem means that quantum key distribution
protocols based on entangled states are also ruled out. At the same
time, single-copy state reconstruction would allow for perfect eavesdropping
on p-states, i.e.\ without leaving a trace.

Collapse-free measurements also modify the computational power of
quantum theory.  Measurement-based quantum computation
\cite{briegel_measurement-based_2009} is evidently impossible in
pQT. In contrast, p-algorithms based on ``quantum parallelism'' turn out to be more powerful than their quantum counterparts since the output of a quantum circuit is ``observable'' via single-copy state reconstruction \cite{kent_quantum_2021, fiorentino_thesis_2025}. 

To see this, let us consider the quantum algorithms by Deutsch and Jozsa, Grover, and Simon
\cite{deutsch_rapid_1992,grover_fast_1996,simon_power_1997}, for
example. They are based on oracles which ``evaluate'' a function $f(x)$
by means of a unitary operator,  $\hat{U}_{f}:\ket{x,0}\to\ket{x,f(x)}$.
Acting with the operator $\hat{U}_{f}$ on the linear combination $\ket s=2^{-\nicefrac{n}{2}}\sum_{x=0}^{2^{n}-1}\ket{x,0}$, the output
state will carry information about all values of the function $f(x)$.
A projective measurement on the state $\hat{U}_{f}\ket s$ will,
however, reveal at most one value of $f(x)$, necessitating further
calls to the oracle.

In the absence of projective measurements, however, all values $f(x)$ can be extracted from the final p-state $\hat{U}_{f}|s\rangle$
by reconstructing it from a \textit{single} copy. Hence, within pQT a single call
to the oracle is sufficient to obtain the result of the computation.  This substantial
reduction in computational cost is accompanied by a large increase in measurement
complexity that is difficult to quantify.

Quantum measurements on their own, i.e.\ without the need for a circuit, allow one to directly access and extract the eigenvalues of any Hermitean matrix  \cite{weigert_quantum_2001}. The protocol only depends on the production of measurement outcomes, not on assigning a specific post-measurement state. Consequently,  p-measurements also possess this particular computational power.  What is more, \textit{all} eigenvalues can be obtained from a single p-state rather than an ensemble as required in the quantum setting.

\subsection{Instruments}
The collapse of a quantum state upon measuring an observable $\hat{A}=\sum_{r}a_{r}\hat{P}_{r}$ has a convenient
description in the language of quantum instruments. The \textit{Lüders} \textit{instrument} consists of a collection of maps $\{\omega_{1}^{\mathsf{L}},\omega_{2}^{\mathsf{L}},\ldots\}$, each sending the initial state $\rho$ to the appropriate (un-normalised) post-measurement state,
\begin{equation}
\rho \, \stackrel{a_{r}}{\longrightarrow} \,\omega_{r}^{\mathsf{L}}\left(\rho\right)=\hat{P}_{r}\rho\hat{P}_{r}\,,\label{eq:Luders instrument}
\end{equation}
conditioned on the outcome $a_{r}$. The trace of the map $\omega_{r}^{\mathsf{L}}$ equals
the outcome probability, $\text{Tr}[\omega_{r}^{\mathsf{L}}\left(\rho\right)]=p_{A}\left(a_{r}\right)$.
Projective measurements act \emph{non-linearly} on the elements $\ket{\psi}$
of Hilbert space $\mathcal{H}$ (cf. Eqs. \eqref{eq: M_3 in QM} and
\eqref{eq: M_3 in QM bipartite}) while the Lüders instrument acts \emph{linearly} on densi-ty
matrices. More generally, \emph{quantum instruments }consist of linear,
completely positive maps \cite{davies_quantum_1976} of density matrices,
all of which can be realised by the Lüders instrument with post-processing
\cite{hayashi_quantum_2017}.

In contrast, pQT is \emph{linear} at the level of p-states, $\ket{\psip}\stackrel{a_{r}}{\to}\ket{\psip}$. The maps $\{\omega_{1}^{\mathsf{P}},\omega_{2}^{\mathsf{P}},\ldots\}$ defining the
associated \textit{p-instrument} act \emph{non-linearly} on density matrices, 
\begin{equation}\label{eq:p-instrument}
\rhop \, \stackrel{a_{r}}{\longrightarrow} \, \omega_{r}^{\mathsf{P}}(\rhop)=\text{Tr}[\hat{P}_{r}\rhop\hat{P}_{r}]\rhop\,.
\end{equation}
Thus, for a pair of density matrices $\rhop_{1}\neq\rhop_{2}$, we find 
\begin{equation}
\omega_{r}^{\mathsf{P}}(\lambda\rhop_{1}+\left(1-\lambda\right)\rhop_{2}) 
\neq
\lambda\omega_{r}^{\mathsf{P}}(\rhop_{1})+\left(1-\lambda\right)\omega_{r}^{\mathsf{P}}(\rhop_{2})\,,
\end{equation}
for $\lambda \in  (0,1)$. This inequality captures the distinguishability of proper and improper mixtures in pQT. Operationally, its left-hand side corresponds to performing a measurement on the \textit{improper} mixture $\rhop=\lambda\rhop_{1}+\left(1-\lambda\right)\rhop_{2}$; the right-hand side describes the effect of a passive measurement on the \textit{proper} mixture of $\rhop_{1}$ and $\rhop_{2}$ with weights $\lambda$ and $\left(1-\lambda\right)$, respectively.

\subsection{Linearity and no-signalling }
A non-linear time evolution of quantum states would, in combination with projective measurements
on entangled states, enable superluminal communication \cite{gisin_stochastic_1989,simon_no-signaling_2001,bassi_no-faster-than-light-signaling_2015-1}.
This result, known as Gisin's argument, is based on the fact that different convex combinations
of quantum states can be used to describe one and the same mixed state.
Using local measurements, an experimenter steers the state of the remote part of a bipartite system into one of two distinct convex combinations corresponding to the same mixed state at the other location. Assuming a non-linear quantum time evolution, a space-like separated observer could subsequently distinguish these decompositions, leading to signalling. However, some fine-tuned non-linear time evolutions cannot be ruled out by Gisin's argument \cite{kent_nonlinearity_2005,ferrero_nonlinear_2004,rembielinski_nonlinear_2020}. In other words, the \emph{linearity }of quantum instruments ensures the peaceful coexistence of the instantaneous, non-local collapse and special relativity.

Alternative state-update rules (rather than alter-native time evolutions) may also result in non-linear transformations of joint and reduced states. Any modification leading to signalling would be unphysical.
In pQT, measurements have no effect on p-states making them entirely unsuitable for signalling. 

\subsection{Simulating quantum theory}
We now show that a physical system realizing pQT can simulate measurements performed on a quantum system, modulo a finite time delay. Recall that after a standard quantum measurement of the observable $\hat{A}$  with (non-degenerate) outcome $a_{r}$, the measured system is known to reside in the associated eigenstate of $\hat{A}$. This is the scenario we wish to replicate. 

First, consider the measurement of a non-degenerate observable $\hat{A}$ on a \textit{single-partite} system in an unknown p-state $|\psip\rangle$. If the outcome $a_{r}$ is obtained, then the experimenter simply replaces the system that continues to reside in the p-state $|\psip\rangle$ by another one residing in the p-state \textit{\emph{$|a_r\rangle$}}. Then, the post-measurement situation is identical to the one in which a quantum measurement with outcome $a_r$  has occurred. An observer without access to the experimenter's lab could not tell apart this procedure from a proper quantum measurement. 

However, substituting the state $|\psip\rangle$ by $|a_r\rangle$ will take a finite amount of time. This ``replacement'' time cannot be made arbitrarily small: if the experimenter generates the state $|a_r\rangle$ from a given state $|\chi\rangle$ by applying an appropriate unitary, ``quantum speed limits'' \cite{deffner_quantum_2017} kick in. Even if the experimenter were to prepare a collection of the eigenstates of the observable $\hat{A}$ in advance, the ``instantaneous'' projection cannot be reproduced since identifying the state $|a_r\rangle$ in the set and physically replacing $|\psip\rangle$ by it will take time.

To reproduce the effect of a local measurement performed on a \textit{bipartite} quantum system $\mathcal{H}_{AB}$ within pQT, the experimenter must, in fact, be able to access all parts of the p-system. Having obtained the (non-degenerate) outcome $a_{r}$ upon measuring the local observable $\hat{A}$ on a known entangled p-state $|\Phi\rangle$, the observer needs to substitute $|\Phi\rangle$ by the appropriate product state $|a_r\rangle\otimes|\psip\rangle$, generated by the quantum mechanical update rule. To identify the correct factor $|\psip\rangle$, the experimenter must reconstruct the p-state $|\Phi\rangle$ through \textit{global} measurements, i.e.\ by using devices implementing operators of the form $\hat{A}\otimes\hat{B}$. Only then will it be possible to produce the correct post-measurement state.

\section{Conclusions} \label{sec: conclusions}
\subsection{Summary \textcolor{modification}{and discussion}}
We have presented a collapse-free foil theory of quantum mechanics characterized by measurements that produce outcomes in line with Born's rule but do not update the state of the measured system. The resulting \textit{passive quantum theory} is a unique tool to investigate the role of the projection postulate. The predictions of pQT agree with those of quantum theory as long as post-measurement states are not used or looked at. Any deviation from the properties of quantum theory must be a consequence of the assumption that states do not collapse when performing measurements in pQT. 

Passive quantum theory does not represent an \emph{interpretation} of quantum theory. Proposals of unitary quantum mechanics, such as modal interpretations \cite{dieks_modal_1998} including Bohm's theory \cite{bohm_suggested_1952} and the relative-state approach \cite{dewitt_many-worlds_2015}, aim to \emph{remove} the need for collapsing quantum states \textit{without} changing the predictions of quantum theory. Many predictions of pQT differ from those of quantum theory.

\textcolor{modification}{Is it logically consistent to modify the standard projection postulate? To do so, it must be independent from the other postulates. 
Ozawa has given a justification of the quantum \textcolor{furthermodification}{instrument formalism for describing state-updates} through Bayes\-ian inference \cite{ozawa_quantum_1984}\textcolor{furthermodification}{, without requiring a separate collapse postulate.} The derivation relies, however, \textcolor{furthermodification}{on using the quantum-mechanical joint probability formula for local measurements, which does \textit{not} hold in pQT. In a sense, the author adopts a \textit{stronger} interpretation of Axiom ${\cal M}_{2}$ (cf.\ Sec.\ \ref{sec: pQT axioms}) than the one used explicitly by us (and implicitly by other authors \cite{kent_nonlinearity_2005, aaronson_space_2016, kent_quantum_2021, kent_measurements_2023} where ${\cal M}_{3}$ is modified). The stronger interpretation of Born's rule is assumed to govern not only individual measurement outcomes but also correlations between \textit{local} measurements carried out by different experimenters with access to parts of a composite system only.} The independence of the quantum mechanical state-update from the other postulates has also been questioned in \cite{masanes_measurement_2019} but continues to be debated \cite{stacey_masanes-galley-muller_2022, kent_measurements_2023} (cf.\ Sec. \ref{Sec: other work})}.

Here are the main differences between pQT and quantum theory we have identified. The possibility of single-copy state reconstruction turns p-states into observable quantities so that any time-evolved state---such as the output of a quantum circuit---can be accessed directly in pQT. Therefore, the number of calls to oracles in well-known quantum algorithms can be reduced considerably; the computational power of these algorithms and their cost must be evaluated anew. Single-copy state reconstruction also implies that passive quantum theory can, in principle, simulate quantum theory \emph{including} the collapse, except for an inevitable finite delay in updating the measured state. In contrast, quantum theory cannot simulate pQT. 

In standard quantum theory, projective measurements can be used to prepare specific states. In pQT, a desired state can only be prepared \textit{dynamically}, i.e.\ by suitably evolving a known state in time.

As is well-known, proper and improper mixtures of quantum states are indistinguishable. Their equiva-lence turns out to be a consequence of the quantum mechanical projection postulate. The collapse-free measurements of pQT, however, allow one to reveal each of the individual states forming a proper mixture. Therefore, these mixtures can be distinguished from improper ones which are still required in order to describe the state of a subsystem. In other words, \textcolor{modification}{in pQT} only improper mixtures are a fundamental concept. 

\textcolor{modification}{I}n quantum theory the observable $\hat{A}\otimes\hat{B}$ can be measured by a single global device $\mathcal{D}_{AB}$, or by two local devices $\mathcal{D}_{A}$ and $\mathcal{D}_{B}$ supplemented by classical communication. The equivalence of the physically different measurement scenarios is a consequence of the way we understand the update rule defined by the projection postulate. In pQT, these two measurement scenarios lead to entirely different outcome statistics.

\subsection{Related work } \label{Sec: other work}
A key feature of passive measurements is the possibility to determine an unknown quantum state without disturbing it. Consequences of this idea have occasionally been addressed in the literature. In his 1997 paper \cite{busch_is_1997}, Busch augments quantum theory by adding a hypothetical procedure to it that allows one to directly \textit{observe} the density operator of an \textit{individual} quantum system. He shows that the proposed  \textit{individual state determination} (ISD) is incompatible with standard quantum theory: it becomes possible to distinguish between proper and improper mixtures and to send superluminal signals when combining ISD with standard, collapse-inducing measurements.

In 2005, hypothetical read-out devices that implement Busch's ISD procedure by providing access to the so-called ``\textit{\emph{local state}}'' of a single system were introduced independently by Kent \cite{kent_nonlinearity_2005, kent_quantum_2021}. Measurements performed with such devices can effectively generate non-linear transformations of states without permitting superluminal signalling, thereby circumventing Gisin's argument \cite{gisin_stochastic_1989}. The \textit{stochastic eigenvalue read-out device} appearing in \cite{kent_quantum_2021} effectively implements passive measurements, and a theory featuring only devices of this type was briefly mentioned in \cite{kent_measurements_2023}.
Read-out devices have also been used to argue that the quantum mechanical state-update rule cannot be derived from the other postulates of quantum theory \cite{kent_measurements_2023}. 

Aaronson et al.\ augment, in 2016, quantum theory by non-disturbing, i.e.\ passive measurements to explore their potential computational advantages \cite{aaronson_space_2016}. They consider a scenario in which passive measurements may be performed at intermediate stages of a quantum circuit, without their output feeding back into the circuit. It turns out that this additional resource does not enable efficient solutions of NP-hard problems. 

Thus, most of the earlier work involving passive measurements can be characterized as \textit{ISD-augmented} quantum theory. The focus of our contribution is, in contrast, on the properties of pQT, an \textit{ISD-only} quantum theory in which non-disturbing measurements \emph{replace} projective measurements rather than coexist with them. 

\subsection{Outlook}
It will be instructive to investigate the consequences of passive measurements more comprehensively, including other quantum information protocols and concepts such as contextuality, POVMs and generalised instruments \cite{fiorentino_pQT_nodate}. It will also be important to clarify the extent to which pQT is a \textit{non-classical} theory---does a description in terms of hidden variables exist? 

Passive measurements clearly represent just one alternative to the projection postulate of quantum theory. Thus, a more general framework is called for, aiming to establish \textit{acceptable state-update rules} based on operational requirements such as the uniqueness of post-measurement states and no-signalling.  In such a family of foil theories \cite{fiorentino_quantum_nodate}, an intriguing question arises: can we identify physical principles which single out the quantum mechanical projection postulate within a larger set of alternatives?






\section*{Acknowledgements}
VF would like to thank the WW Smith Fund at the University of York for financial support. This work was partially supported by the Leverhulme Trust, grant number RPG-2024-201.





\end{document}